# Calculations of the Decay Transitions of the Modified Pöschl-Teller Potential Model via Bohr Hamiltonian Technique

Nahid Soheibi[1] , Majid Hamzavi[1], Mahdi Eshghi[2,*], Sameer M. Ikhdair[3,4]

[1] *Department of Physics, University of Zanjan, Zanjan, Iran*

[2] *Yang Researchers and Elite Club, Central Tehran Branch, Islamic Azad University, Tehran, Iran*

[3] *Department of Physics, Faculty of Science, an-Najah National University, Nablus, Palestine*

[4] *Department of Electrical Engineering, Near East University, Nicosia, Northern Cyprus, Mersin 10, Turkey*

**Abstract**

We calculate the eigenvalues and their corresponding eigenfunctions of the Bohr's collective Hamiltonian with the help of the modified Pöschl-Teller potential model within $\gamma$-unstable structure. Our numerical results for the ground state β and γ band heads together with the electric quadrupole B(E2) transition rates are displayed and compared with those available experimental data.

## 1. Introduction

[*] *Corresponding Author Email: eshgi54@gmail.com*

The Bohr Hamiltonian [1] and model extension along with the geometrical collective model [2, 3], have been provided, for several decades, which used as a phenomenological framework in understanding the collective behavior of atomic nuclei. The shape phase transitions in nuclei are currently under main study in both theoretical and experimental perspectives. Recently, a new phase study started with the shape phase transitions by using the classical limits of Hamiltonians and Lie algebras [4-6].

On the other hand, dynamic symmetries have provided a useful tool to describe properties of several physical systems. One essential important example appears in nuclear physics is the dynamic symmetries of the interacting boson model [7] and another example in molecular physics is Vibron model [8]. Furthermore, the application of this topic is seen in nature of the quantum phase transition [9] between the dynamical symmetries in nuclei: such as U(5), SU(3), SO(6) into an interacting boson model. Also, the X(5) symmetry [10] is designed to describe the first-order phase transition between vibrational and axially symmetries prolate deformed rotational nuclei, In fact, this symmetry corresponds to transition from vibrational spherical shape U(5) to prolate deformed nuclei S(3), and the E(5) [11]. The symmetry is designed to describe the second-order phase transition between spherical and $\gamma$-unstable nuclei, namely, this symmetry corresponds to transition from vibrational U(5) to $\gamma$-unstable nuclei. Actually, the symmetry E(5) represents a solution of the Bohr Hamiltonian with a $\gamma$-independent potential while the symmetry X(5) has a minimum at $\gamma = 0$ for the case of the $\gamma$ potential. In all these cases, dynamical symmetries are related to exactly solvable problems and also produce results for observables in explicit analytic form. Recently, with the aim of describing appropriately shape phase transitions in atomic nuclei, some authors have devoted to construct analytical solutions of the Schrodinger equation associated with the Bohr Hamiltonian using various potential models [12-18].

In the present work, we study the modified Pöschl-Teller (MPT) potential model which is a short-range model potential. This potential model has been used to describe bending? molecular vibrations [19−21]. Therefore, with the modified Pöschl-Teller potential, we calculate the energies and $B(E2)$ transition rates of Bohr Hamiltonian by using the Nikiforov-Uvarov (NU) method.

This paper is organized as follows: In Section 2, we present, in brief, the model used in our work. In Section 3, we calculate the energies of the MPT potential model and electric quadrupole $B(E2)$ transition rates. In Section 4, we give our concluding remarks.

## 2. Bohr Hamiltonian Theory

### *2.1. Energy spectrum and wavfunctions*

We present the Schrodinger equation in the form [22]

$$H\psi_{nlm}(\beta,\gamma,\theta_i) = E\psi_{nlm}(\beta,\gamma,\theta_i), \tag{1}$$

where $\psi_{nlm}(\beta,\gamma,\theta_i)$ is the wave function, $E$ and $H$ are energy eigenvalues and Hamiltonian, respectively. Further, the Bohr Hamiltonian is well defined by [2]

$$H=-\frac{\hbar^2}{2B}\left[\frac{1}{\beta^4}\frac{\partial}{\partial\beta}\beta^4\frac{\partial}{\partial\beta}+\frac{1}{\beta^2\sin3\gamma}\frac{\partial}{\partial\gamma}\sin3\gamma\frac{\partial}{\partial\gamma}\right.$$

$$\left.-\frac{1}{4\beta^2}\sum\frac{Q_k^2}{\sin^2(\gamma-\frac{2}{3}\pi k)}\right]+V(\beta,\gamma), \tag{2}$$

where $Q_k$ $(k=1,2,3)$ are the components of angular momentum in the intrinsic frame, and B? is the mass parameter and the potential form is $V(\beta,\gamma)=U(\beta)+\frac{1}{\beta^2}U'(\gamma)$. Also $\beta$ and $\gamma$ are the usual collective coordinates [23, 24].

The original wave function is defined in the form

$$\psi_{nlm}(\beta,\gamma,\theta_i)=\xi(\beta)\phi(\gamma,\theta_i), \tag{3}$$

where $\theta_i (i=1,2,3)$ are the Euler angles. In addition, the potential is mainly dependent on $\beta$ variable for $\gamma$-unstable structures. Now, we need to solve Eqs. (1) and (2) for its eigenvalues. After making a separation of variables, one obtains

$$\left[-\frac{1}{\beta^4}\frac{\partial}{\partial\beta}\beta^4\frac{\partial}{\partial\beta}+u(\beta)+\frac{\Lambda}{\beta^2}\right]\xi(\beta)=\varepsilon\xi(\beta), \tag{4a}$$

$$\left[-\frac{1}{\sin 3\gamma}\frac{\partial}{\partial\gamma}\sin 3\gamma\frac{\partial}{\partial\gamma}+\frac{1}{4}\sum\frac{Q_k^2}{\sin^2(\gamma-\frac{2}{3}\pi k)}+u'(\gamma)\right]\phi(\gamma,\theta_i)=\Lambda\phi(\gamma,\theta_i).$$

(4b)

where we have used $u=\frac{2B}{\hbar^2}U, u'=\frac{2B}{\hbar^2}U', \varepsilon=\frac{2B}{\hbar^2}E$

In the above equation, $\Lambda$ is a separation constant usually being expressed as pointed out in ref.[17] as $\Lambda=\tau(\tau+3)=9n_\gamma(n_\gamma+1)+3\sqrt{c+s}(2n_\gamma+1)+c+\frac{L(L+4)+3n_\omega(2L-n_\omega)}{4}$ and $\tau\,(\tau=0,1,2,...)$ is the seniority quantum number [25,17]. Likewise $n_\omega, n_\gamma$ distinctive as wobbling quantum number and other quantum number that related to $\gamma$–excitation, respectively. In this equation c and s are free parameters.

In solving the radial part equation given by Eq. (4a), we use the modified Pöschl-Teller (MPT) potential [22] in the form:

$$u(\beta)=-\frac{1}{\cosh^2(\alpha\beta)}, \tag{5}$$

where $\alpha$ is the range of the MPT potential. After inserting Eq. (5) into Eq. (4a), we can obtain

$$\left[-\frac{\partial^2}{\partial\beta^2}+\frac{\Lambda}{\beta^2}-\frac{4}{\beta}\frac{\partial}{\partial\beta}-\frac{1}{\cosh^2(\alpha\beta)}\right]\xi(\beta)=\varepsilon\xi(\beta). \tag{6}$$

Further, we need to make the following conversion of parameters $\xi(\beta)=\beta^2\chi(\beta)$ so as to establish the radial wave function as

$$\left[-\frac{\partial^2}{\partial\beta^2}+\frac{\Lambda+2}{\beta^2}-\frac{1}{\cosh^2(\alpha\beta)}\right]\chi(\beta)=\varepsilon\chi(\beta). \tag{7}$$

The solution of the above radial part is possible when using an approximation to the term $\frac{1}{\beta^2}$ as [22]

$$\frac{1}{\beta^2}=\alpha^2\left[4d_0+\frac{1}{\sinh^2(\alpha\beta)}\right], \tag{8}$$

where the constant $d_0=\frac{1}{12}$. Thus, Eq. (7) simplifies in the form as

$$\frac{\partial^2\chi}{\partial\beta^2}+\left[\varepsilon+\frac{1}{\cosh^2(\alpha\beta)}-\frac{e}{3}-\frac{e}{\sinh^2(\alpha\beta)}\right]\chi=0. \tag{9}$$

where $e=(\Lambda+2)\alpha^2$. The last equation can be solved by defining new variable as $z=\tanh^2(\alpha\beta)$. Therefore, the above equation (9) reduces to the quite simple form

$$\frac{d^2\chi}{dz^2}+\frac{(\frac{1}{2}-\frac{3}{2}z)}{z(1-z)}\frac{d\chi}{dz}+\frac{1}{4\alpha^2z^2(1-z)^2}\left[-z^2+z(-\varepsilon+1+\frac{4}{3}e)-e\right]\chi=0$$

(10)

Now we use the parametric generalization of the Nikiforov-Uvarov (pNU) method [25-26] to solve the above equation. This method is usually used to solve such a second-order differential equation [25-26]. The energy states can be obtained as

$$\varepsilon=\frac{e}{3}-4\alpha^2\left[\sqrt{\frac{1}{16}+\frac{1}{4\alpha^2}}-\sqrt{\frac{1}{16}+\frac{e}{4\alpha^2}}-n-\frac{1}{2}\right]^2, \tag{11}$$

where $n$ is the principal quantum number. According to this method, the wave functions are given by

$$\psi_{nk} = N_{nk} z^{(\frac{1}{4}+\sqrt{\frac{1}{16}+\frac{\epsilon}{4\alpha^2}})}(1-z)^{\left|-n-\frac{1}{2}-\sqrt{\frac{1}{16}+\frac{\epsilon}{4\alpha^2}}+\sqrt{\frac{1}{16}+\frac{1}{4\alpha^2}}\right|} \times$$

$$P_n^{\left(2\sqrt{\frac{1}{16}+\frac{\epsilon}{4\alpha^2}},\, 2\left|-n-\frac{1}{2}\sqrt{\frac{1}{16}+\frac{\epsilon}{4\alpha^2}}+\sqrt{\frac{1}{16}+\frac{1}{4\alpha^2}}\right|\right)}(1-2z). \tag{12}$$

In above equation $P_n^{(\mu,\nu)}(x),\ \mu>-1,\ \nu>-1,$ and $x\in[-1,1]$ are Jacobi Polynomials and $N_{nk}$ is a normalization constant.

### *2.2. B(E2) transition rates*

Having found the expression of the total wave function, it leads to calculate the B(E2) transition rates [27]. The B(E2) transition rates from an initial to a final state are defined as [17,27]

$$B(E2;L_i\alpha_i \to L_f\alpha_f) = \frac{5}{16\pi}\frac{\left|\langle L_f\alpha_f \| T^{(E2)} \| L_i\alpha_i\rangle\right|^2}{(2L_i+1)}, \tag{13}$$

where the reduced matrix elements are being calculated by Wigner-Eckrat theorm [27]

$$\langle L_f M_f \alpha_f | T_M^{(E2)} | L_i M_i \alpha_i\rangle = \frac{(L_i 2 L_f | M_i M M_f)}{\sqrt{2L_f+1}}\langle L_f\alpha_f \| T^{(E2)} \| L_i\alpha_i\rangle. \tag{14}$$

While M is the quantum numbers of the projections of angular momentum on the laboratory fixed $z$ -axis and $\alpha$ is the quantum numbers of the projections of angular momentum on the body-fixed $\acute{x}$-axis, respectively. In our calculation to the matrix elements of the quadrupole operator from Eq. (14), the integral over the Euler angles is implemented by means of the standard integrals of three Wigner functions [27], while the integral over β is given via the expression:

$$I_\beta(n_i,L_i,\alpha_i,n_f,L_f,\alpha_f) = \int_0^\infty \beta\, \xi_{n_i,L_i,\alpha_i}(\beta)\, \xi_{n_f,L_f,\alpha f}(\beta)\, \beta^4 d\beta. \tag{15}$$

## 3. Results and Discussions

In the present work, our theoretical predictions for the energy levels given via Eq. (11) are evaluated. This equation depends on three parameters, the screening parameter $\alpha$ in the $\beta$ potential and the rig-shape parameters c and s of the $\gamma$ potential [17]. To obtain the potential parameters for each nuclei, we need to evaluate the root mean square (rms) deviation between the experimental data and the theoretical ones by means of the equation

$$\sigma=\sqrt{\frac{\sum_{i=1}^{m}(E_i(\exp)-E_i(th))^2}{(m-1)E(2_1^+)^2}}, \qquad (16)$$

where $m$ mentions the number of states, while $E_i(th)$ and $E_i(\exp)$ represent the theoretical and experimental energy of the $i-th$ level, respectively. We know $E(2_1^+)$ as the energy of the first excited level of the ground state band [17]. The calculated results are displayed in Table 1 for each nucleus. We further demonstrate the comparison of theoretical predictions of the γ-unstable to experimental data for the ground state (g.s.) band head, β and γ band heads, normalized to the $E(2^+_{g.s.})$ state and labeled by $R_{4/2}=E(4^+_{g.s.})/E(2^+_{g.s.})$, $R_{0/2}=E(0^+_{\beta})/E(2^+_{g.s.})$ and $R_{2/2}=E(2^+_{\gamma})/E(2^+_{g.s.})$ ratios respectively. For Xe and Pt the calculated ratios are not very different from experimental data and a lot of them have good agreement with each other's. As shown in Table 1, deviation from experimental data is negligible. The isotopic dependence of $R_{4/2}$ versus mass number for Xe and Pt nucleus are plotted in Figs. 1, 2 [28]. The present calculations provide obvious decrease in $R_{4/2}$ with mass number. We also find a good agreement in the comparison of experimental data and theoretical ones. This agreement with the experimental ratios for Xe isotopes (see Figure 1) is better than for Pt isotopes (see Figure 2). The largest disagreement between theoretical and experimental ratios for Xe is about 0.05. But for Pt the major difference is about 0.07 and there is no any close agreement in plotted points. In Table 2, we compare the $B(E2)$ transition rates with their corresponding experimental data for the same isotopes as in Table 1. The transition probabilities for electric quadrupole radiation are normalized to the experimental $B(E2; 2^+_{g.s} \rightarrow 0^+_{g.s})$. All bands (ground state, $\beta$ and $\gamma$) are introduced by the quantum numbers, $n$ , $n_w$ , $n_\gamma$ and L. The ground state band is characterized by $n = 0,\ n_\gamma = 0,\ n_w = 0$ and the $\beta$ band is characterized by $n = 1,\ n_\gamma = 0,\ n_w = 0$. Notice that the $\gamma$ band composed by the even and odd levels with $n = 0,\ n_\gamma = 0,\ n_w = 2$ and $n = 0,\ n_\gamma = 0,\ n_w = 1$, respectively. The obtained results in Table 2 are not satisfactory. (Remove them?)However, there are small deviations between the predicted and experimental values. For example, in the case of $^{128}$Xe isotope, the agreement is somewhat better than in other isotopes. Finally, to have a better view of the results, we can depict the energy levels and also $B(E2)$ transitions for the same isotopes in Figures 3 to 10. Moreover, one can easily compare the theoretical and experimental results in these Figures. According to Table 2 and Figures 3 to 10, the excitation spectrum of the collective Hamiltonian is in good agreement with the available data for the $^{128}$Xe in the excitation energies and transitions in the ground-state band. But for other isotopes and nuclei, the excitation energies and transitions in the $\beta$ and $\gamma$ bands are nearly better than ground-state band. For example, for $^{196}$Pt the excitation energies and transitions in the $\beta$ band are in close agreement with corresponding experimental data.

## 4. Conclusions

In this paper, by means of the generalized parametric NU method, we have solved the Bohr Hamiltonian in the framework of the modified Pöschl-Teller potential model. To show the accuracy of the present theoretical results, we have obtained some numerical calculations for the energy spectra as well as for the electric quadrupole $B(E2)$ transitions. It can be seen that our results are in good agreement with the experimental ones.

Acknowledgements: The authors would like to thank the kind referee(s) for the positive suggestions and critical reading of the text which have greatly improved the present text.

## References

1. Bohr A. Mat. Fys. Medd. K. Dan. Vidensk. Selsk. **26** (1952) 14.
2. Bohr A. and Mettelson B.R. "*Nuclear Structure*", Vol. II: Nuclear Deformations, Benjamin, New York, 1975.
3. Eisenberg J.M. and Greiner W. "*Nuclear Theory",* Vol. I: Nuclear Models, North-Holland, Amsterdam. 1975.
4. Gilmore R. *J. Math. Phys*. **20** (1979) 891.
5. Gilmore R. and Feng D.H. *Phys. Lett. B* **76** (1978) 26.
6. Gilmore R. and Feng D.H. *Nucl. Phys. A* **301** (1978) 189.
7. Iachello F. and Arima A. "*The interacting Boson Model".* Cambridge University Press, Cambridge, England, 1995.

8. Iachello F. and Levine R. “*Algebraic Theory of Molecules”*. Oxford University Press, Oxford, England, 1995.
9. Dieperink A.E.L., Scholten O. and Iachello F. *Phys. Rev. Lett.* **44** (1980) 1747.
10. Iachello F. *Phys. Rev. Lett*. **87** (2001) 052502.
11. Iachello F. *Phys. Rev. Lett*. **85** (2000) 3580.
12. Iachello F. *Phys. Rev. Lett*. **91** (2003) 132502.
13. Fortunato L. and Vitturi A. *J. Phys. G: Nucl. Part. Phys*. **29** (2003) 1341.
14. Cejnar P., Joli J. and Casten R.F. *Rev. Mod. Phys*. **82** (2010) 2155.
15. Bonatsos D., Georgoudis P.E., Minkov N., Petrellis D. and Quesne C. *Phys. Rev. C* **88** (2013) 034316.
16. Chabab M., Lahbas A. and Oulne M. *Int. J. Mod. Phys. E* **24**(11) (2015) 1550089.
17. Chabab M., Lahbas A. and Oulne M. *Eur. Phys. J. A* **51** (2015) 131.
18. Chabab M., Batoul A. El, Lahbas A.and Oulne M.; *Phys. Lett. B* **758** (2016) 212.
19. Iachello F. and Oss S. *Chem. Phys. Lett*. **205** (1993) 285.
20. Iachello F. and Oss S. *J. Chem. Phys*. **99** (1993) 7337.
21. Agboola D. *Chin. Phys. Lett.* **27** (2010) 040301
22. Chabab M., Batoul A. El, Hamzavi M., Lahbas A. and Oulne M., *Eur. Phys. J. A* **53** (2017) 157.
23. Chabab M., Batoul A. El, Lahbas A. and Oulne M. *Nucl. Phys. A* **953** (2016) 158
24. Rakavy G., *Nucl. Phys*. **4** (1957) 289
25. Nikiforov A. F. and Uvarov V. B., *Special Functions of Mathematical Physics*, Birkhausr, Berlin, 1988.
26. Amirfakhrian M., Hamzavi M., *Mol. Phys*. **110** (2012) 2173.

27. Edmonds A. R., *Angular Momentum in Quantum Mechanics*, Princeton University Press, Princeton, 1957.
28. Li Z.P., Niksic T., Vretenar D. and Meng J., *Phys. Rev. C* **81** (2010) 034316.

**Table 1.** Comparison of the theoretical predictions of the γ-unstable with the experimental data for the ground state β and γ band heads, normalized to the $E(2^+_{g.s.})$ state and labeled by $R_{4/2}=E(4^+_{g.s.})/E(2^+_{g.s.})$, $R_{0/2}=E(0^+_{\beta})/E(2^+_{g.s.})$ and $R_{2/2}=E(2^+_{\gamma})/E(2^+_{g.s.})$ ratios, respectively.

| $Nucl.$ | $R_{4/2}$ Exp | $R_{4/2}$ Th | $R_{0/2}$ Exp | $R_{0/2}$ Th | $R_{2/2}$ Exp | $R_{2/2}$ Th | $rms$ | $\alpha$ | C | S |
|---|---|---|---|---|---|---|---|---|---|---|
| $^{192}Pt$ | 2.479 | 2.42916 | 3.776 | 3.89833 | 1.935 | 1.88744 | 0.52 | 0.083 | 2.083 | 7.936 |
| $^{194}Pt$ | 2.470 | 2.41766 | 3.858 | 3.86047 | 1.894 | 1.88153 | 0.54 | 0.079 | 2.151 | 5.114 |
| $^{196}Pt$ | 2.465 | 2.39356 | 3.192 | 3.40402 | 1.936 | 1.86997 | 0.64 | 0.099 | 3.476 | 1.250 |
| $^{126}Xe$ | 2.424 | 2.47247 | 3.381 | 3.63966 | 2.264 | 1.90944 | 0.72 | 0.101 | 6.172 | 15.995 |
| $^{128}Xe$ | 2.333 | 2.32892 | 3.574 | 3.47425 | 2.189 | 1.83365 | 0.49 | 0.014 | 0.490 | 0.489 |
| $^{130}Xe$ | 2.247 | 2.26392 | 3.346 | 3.06256 | 2.093 | 1.79849 | 0.29 | 0.015 | 0.042 | 0.329 |
| $^{132}Xe$ | 2.157 | 2.10227 | 2.771 | 2.25892 | 1.944 | 1.71093 | 0.37 | 0.087 | 0.007 | 0.0002 |
| $^{134}Xe$ | 2.044 | 2.09865 | 1.932 | 2.2225 | 1.905 | 1.70961 | 0.60 | 0.098 | 0.006 | 0.009 |

**Table 2.** The electric quadrupole $B(E2)$ transition rates

| Nucl. | | $\frac{4_g \to 2_g}{2_g \to 0_g}$ | $\frac{6_g \to 4_g}{2_g \to 0_g}$ | $\frac{8_g \to 6_g}{2_g \to 0_g}$ | $\frac{10_g \to 8_g}{2_g \to 0_g}$ | $\frac{0_\beta \to 2_g}{2_g \to 0_g}$ | $\frac{2_\beta \to 0_g}{2_g \to 0_g} \times 10^3$ | $\frac{2_\gamma \to 2_g}{2_g \to 0_g}$ | $\frac{2_\gamma \to 0_g}{2_g \to 0_g} \times 10^3$ |
|---|---|---|---|---|---|---|---|---|---|
| $^{192}Pt$ | Exp | 1.56 | 1.23 | ….. | ….. | ….. | ….. | 1.91 | 9.5 |
| | Theor | 1.798 | 2.864 | 4.649 | 8.136 | 0.065 | 0.072 | 1.642 | 1.150 |
| $^{194}Pt$ | Exp | 1.73 | 1.36 | 1.02 | 0.69 | 0.01 | ….. | 1.81 | 5.9 |
| | Theor | 1.769 | 2.743 | 4.272 | 7.017 | 0.064 | 0.071 | 1.627 | 1.139 |
| $^{196}Pt$ | Exp | 1.48 | 1.80 | 1.92 | ….. | 0.07 | 0.06 | ….. | 0.4 |
| | Theor | 1.987 | 3.680 | 7.531 | 19.387 | 0.073 | 0.088 | 1.742 | 1.220 |
| $^{128}Xe$ | Exp | 1.47 | 1.94 | 2.39 | 2.74 | ….. | ….. | 1.19 | 15.9 |
| | Theor | 1.473 | 1.780 | 2.017 | 2.219 | 0.062 | 0.056 | 1.456 | 1.019 |
| $^{132}Xe$ | Exp | 1.24 | ….. | …… | ….. | ….. | ….. | 1.77 | 3.4 |
| | Theor | 1.949 | 3.296 | 5.588 | 10.081 | 0.083 | 0.091 | 1.734 | 1.213 |

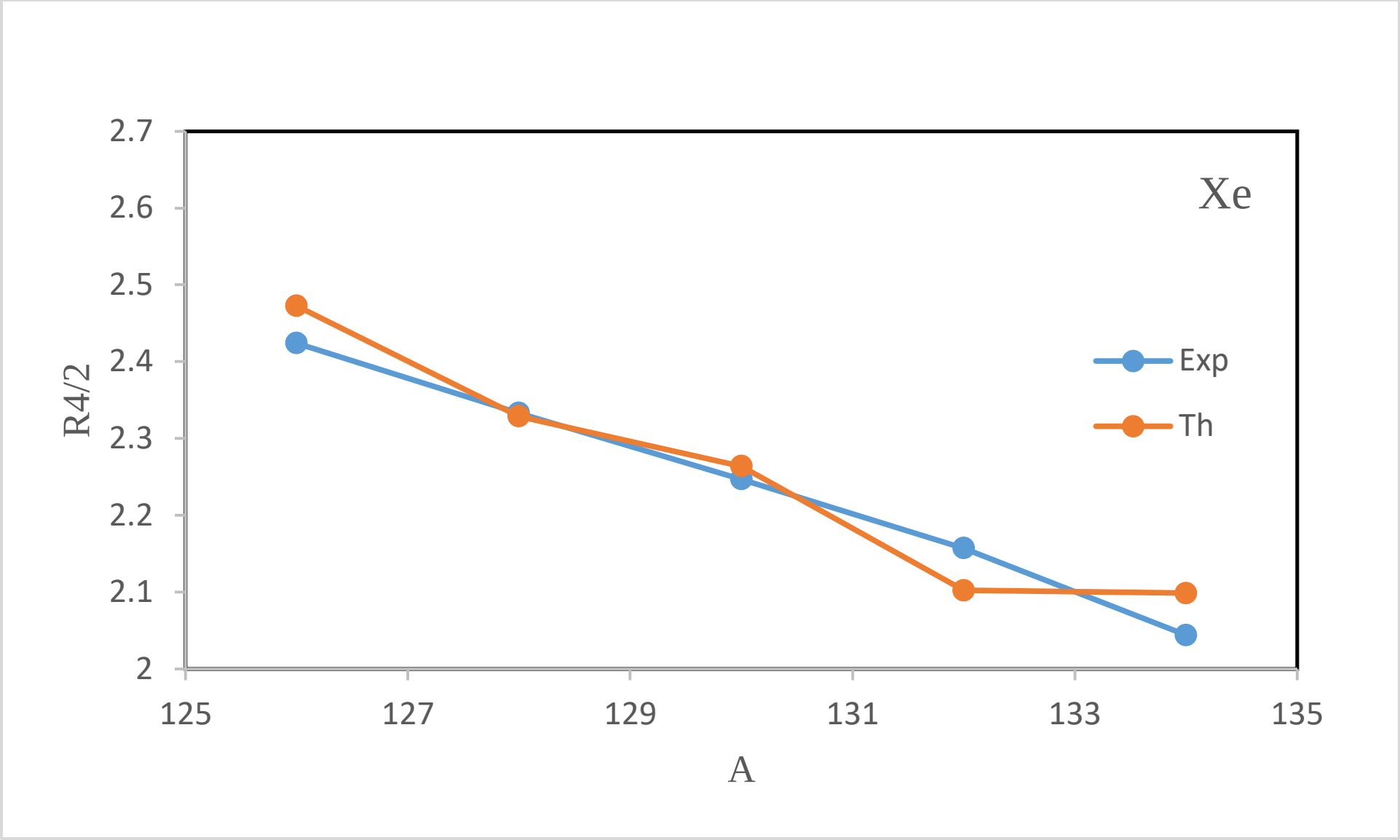

**Figure 1:** Evolution of the $R_{4/2}$ with mass number in Xe isotopes.

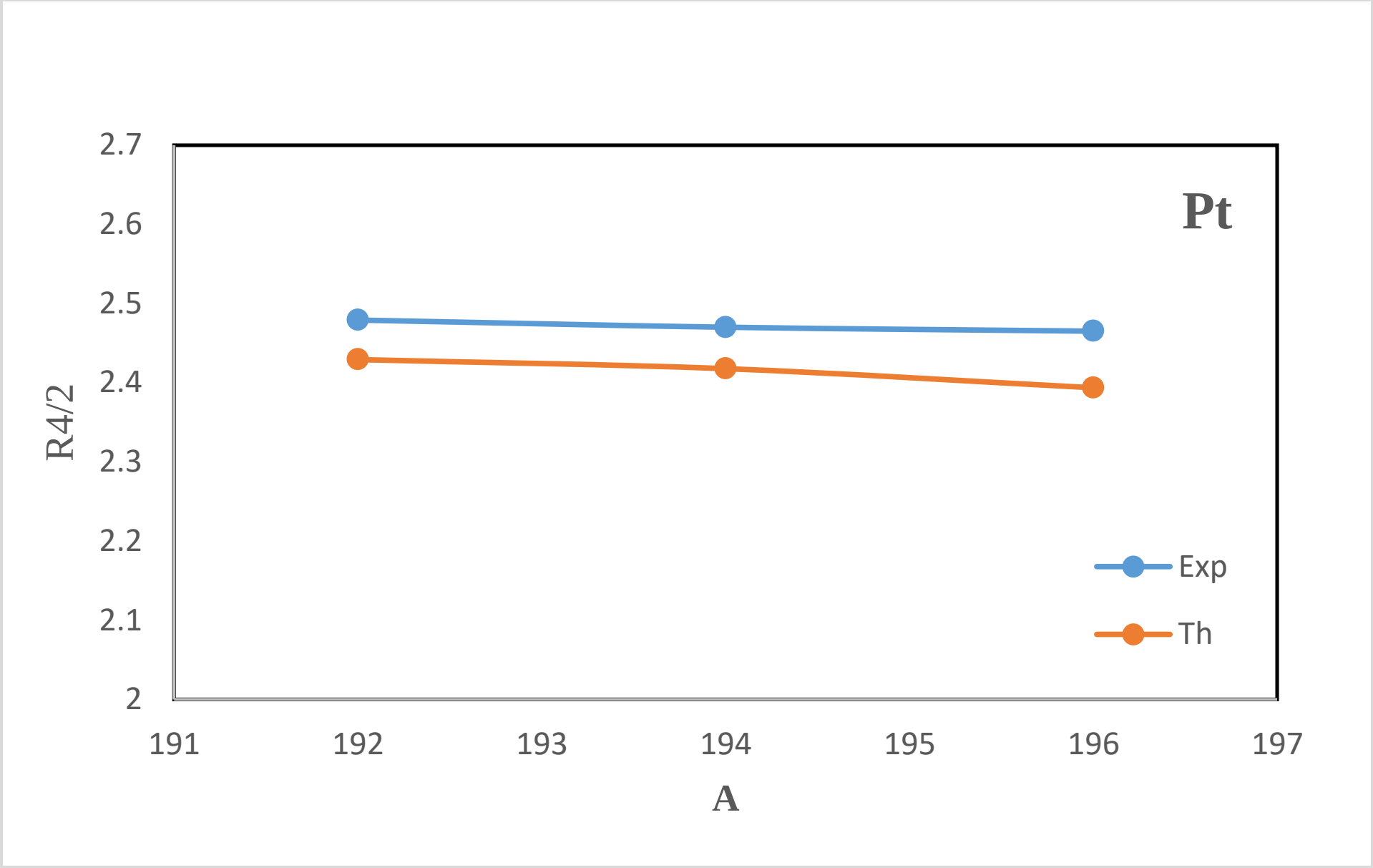

**Figure 2:** Evolution of the $R_{4/2}$ with mass number in Pt isotopes.

**Figure 3:** Comparison of the theoretical energy spectra and the experimental ones for $^{126}$Xe

**Fig.4.** Same as Fig. 3 for $^{128}$Xe

**Fig.5.** Same as Fig. 3 for $^{130}$Xe

**Fig.6.** Same as Fig. 3 for $^{132}$Xe

**Fig.7.** Same as Fig. 3 for $^{134}$Xe

**Fig.8.** Same as Fig. 3 for $^{192}$Pt

**Fig.9.** Same as Fig. 3 for $^{194}$Pt

**Fig.10.** Same as Fig. 3 for $^{196}$Pt